# Minority Game of price promotions in fast moving consumer goods markets

Robert D. Groot[*] and Pieter A. D. Musters

Unilever Research Vlaardingen

PO Box 114, 3130 AC Vlaardingen, The Netherlands

A variation of the Minority Game has been applied to study the timing of promotional actions at retailers in the fast moving consumer goods market. The underlying hypotheses for this work are that price promotions are more effective when fewer than average competitors do a promotion, and that a promotion strategy can be based on past sales data. The first assumption has been checked by analysing 1467 promotional actions for three products on the Dutch market (ketchup, mayonnaise and curry sauce) over a 120-week period, both on an aggregated level and on retailer chain level.

The second assumption was tested by analysing past sales data with the Minority Game. This revealed that high or low competitor promotional pressure for actual ketchup, mayonnaise, curry sauce and barbecue sauce markets is to some extent predictable up to a forecast of some 10 weeks. Whereas a random guess would be right 50% of the time, a single-agent game can predict the market with a success rate of 56% for a 6 to 9 week forecast. This number is *the same* for all four mentioned fast moving consumer markets. For a multi-agent game a larger variability in the success rate is obtained, but predictability can be as high as 65%.

Contrary to expectation, the actual market does the *opposite* of what game theory would predict. This points at a systematic oscillation in the market. Even though this result is not fully understood, merely observing that this trend is present in the data could lead to exploitable trading benefits. As a check, random history strings were generated from which the statistical variation in the game prediction was studied. This shows that the odds are 1:1,000,000 that the observed pattern in the market is based on coincidence.

# 1. Introduction

The fast moving consumer goods (FMCG) market covers the daily needs of consumers, such as foods and detergents, bought at retailer outlets. Every brand tries to draw the attention of the consumer by advertisements and price-promotions. As a result, the market is characterised by strong sales peaks that are caused by promotional actions. Promotions are typically initiated by brand managers, in co-operation with retailers, and have to be planned some eight weeks in advance in view of the logistics of extra factory production, preparation of display material etc.

The problem for a brand manager is to decide on the form of an action, and on its timing. Even if the form of the action is kept constant, it is envisaged that it will be advantageous to plan a promotion when fewer than average competitors have a promotion. The central question of this paper is therefore, is it possible to plan promotional actions via a strategy, such that fewer than average competitors will plan an action in that particular week?

Since the result of any strategy to plan promotion timings depends on the actions of all competing brand managers, it is impossible to derive the winning strategy from first principles. The brand managers simply lack information on what the others will do. Brian Arthur recently suggested that when problems become too complicated, humans use rules of thumb that worked in the past, rather than trying to calculate the best strategy in a rational way [1]. This seems to be more in line with psychological models of how humans reach decisions than brute force calculation is. In this famous the El Farol Bar problem, Arthur proposed a simulation scheme to model this process, where agents "learn" which of their hypotheses work, and from time to time discard poorly performing hypotheses and generate new "ideas" to put in their place. A slightly simplified version of this problem is the Minority Game [2,3,4,5], in which agents choose between two options (0 or 1), and the agents who are in the minority win. This game has been solved exactly by Marsili *et al* [6].

The relevance of this game to marketing problems is that brand managers are competing for market share by adapting their price and advertisement on the basis of a strategy, and above all they seek the difference from their competitors. That is, they try to be in the minority. Since they have no information about the strategies of the other brand managers, there is no obvious "best" strategy. If there was one, all brand managers would follow that, and no one would win — negating the efficacy of this strategy. Indeed, the Minority Game has been applied successfully to model stock markets [7,8]. This demonstrated that the Minority Game reproduces aspects of real markets such as clustered volatility and broad distributions in price fluctuations, but also it shows an interplay between exploitable predictability, price fluctuation and trader behaviour [9].

One key result of the Minority Game is market predictability. In brief, markets can operate in two distinct modes, or phases. These correspond to an equilibrium market, and an out-of-equilibrium phase where demand and supply do not match. Theory predicts that a real market will always operate close to the edge between these two phases [10,11]. This is indeed observed in real financial markets [12]. The important point now is that when markets are in the equilibrium phase, there are no market opportunities and the market is unpredictable. The reverse is true when markets are out of equilibrium: there are market opportunities (e.g. because there is an excess demand) and as a consequence of this the market becomes (slightly) predictable.

This predictability has been demonstrated by Johnson *et al* on real financial time series [13] and by Lamper *et al* for a simulated market [14]. In particular, Johnson *et al* investigated if the Minority Game can be used to predict whether the exchange rate of Dollars against Yens will go up or down in the next hour, based on the past hourly exchange rate history. In contrast to a random guess that would be right 50% of the time, they found that the game could predict the market correctly 54% of the time.

In this paper we investigate if market predictability also holds for fast moving consumer goods markets. Particularly, can we apply the Minority Game to improve marketing strategies? To predict optimal promotion timings, we need to make an eight-week forecast of what competitors will do, as this is the time lag between planning with the retailer and actual promotion. To our knowledge this has not been done. In the next section we will give a brief introduction of the market studied, followed by a description of the games used to analyse the market data. The results of this analysis are presented in section 4, and conclusions are drawn in section 5.

## 2. Sales in fast moving consumer goods markets

As an example of sales in a FMCG market, promotional sales data of ketchup in the Netherlands will be discussed first, and then other commodities will be included. Ideally we would like to know how many promotional actions are being run simultaneously in the environment of a consumer, but as a first guess we may assume that this is proportional to sales under promotion. Time series of sales were obtained from AC-Nielsen [15]. The longest series available is over a 120-week period. AC-Nielsen standard gives the sales per brand, but separately gives the sales in shops where a promotion is on (a price discount which can be supported by a display) and in shops without promotion activity. The lower curve in Figure 1 gives the sales under promotion, the upper curve gives the total ketchup sales in the Netherlands, in number of items sold. The data reported is actual point of sales data covering some 5000 retail shops, but it is based on a representative sample of about 350 shops.

The underlying assumption of the present work is that promotions are more profitable if fewer competitors have a promotion in the same week. Is this actually correct? To investigate this, we checked the database for overlap between promotional actions in the same week within the same retailer chain and across retailers over the 120-week period available. In most cases promotional sale has a big peak (typically $10^4$ items sold) in the first week, followed by a tail of 500 in the second week and 200 in the third. If we disregard this tail, we are left over with 91 major events. Of these, 16 promotions overlapped with another action within the same retailer. If we plot the promotional sales of one brand as function of the simultaneous promotional sales of all other brands within the same retailer AH (eight events) we find a straight line of slope –0.96±0.12. This suggests that the gain of one action is taken away from the other. However, from the available data we cannot exclude that two actions run simultaneously within the same chain, but in different places.

If we plot all ketchup promotions within AH as function of promotional ketchup sales at any other retailer, 33 points follow a slope of –0.38±0.12. Similar numbers are found for other retailers, but the number of events is too small to draw a definitive conclusion. An explanation could be that consumers visit several shops per week, and buy an item that is on promotion in one shop that they would otherwise have bought elsewhere. These could be deal-to-deal buyers. Indeed, deal-to-deal buying and brand switching are generally seen to contribute about 1/3 each of the promotional sales [16]. The advantage of being the only one to do a promotion has thus been demonstrated qualitatively within individual retailers and across retailers.

To get a more quantitative estimate of the mutual influence of promotions, we need to make an average over many promotion events. Is it justified to average over brands and products? To check this, all ketchup promotion sales of any brand, anywhere in the Netherlands over a 120-week period are shown in Figure 2a on a log-log scale. Even though this shows a very large scatter, a power law fit shows a declining trend towards the high competitor promotion side. Moreover, all four brands shown follow the same statistical pattern. For this reason we are confident that we can average over the various brands, to arrive at a ketchup market average trend. To this end all points are sorted to their x-co-ordinate, and averages are taken over each successive 30 points.

These averages are shown in the in Figure 2b by the crosses. Similar analyses for the mayonnaise and curry markets are shown by the open and closed triangles respectively. All three markets show a very similar trend, so that we can indeed take all 1467 promotional actions of ketchup, mayonnaise and curry sauce together. Averages over 100 successive promotions are shown by the black dots that fall on a straight line of decreasing slope. If we take out Heinz and Retailer Own Brands from the statistics, we obtain the black circles that show the same trend. This data indicates that the same trend holds across all brands and products studied. Figure 2b also shows a dashed vertical line. This line indicates the x-co-ordinate to which 50% of all promotions lie to the left, and 50% to the right. The left 50% are the promotions that satisfy the criterion "conducted when the competitor action is less than average" whereas the right 50% do not satisfy this criterion.

## 3. The Minority Game

To pick out a favourable day for a promotion we need to predict where the next peak will be, based on the sales history of the past few weeks. This comes down to finding patterns in the data. If all actions are timed randomly they are unpredictable, but if they are based on a promotion strategy that takes other actions into account, the promotional sales will show subtle predictive patterns. If predictive patterns are present in the time series, they can be found using game theory, because in this method we search for all possible patterns. To find such predictable patterns, the data of sales under promotion σ(t) is first transferred into a bit string. For every week the sales data is compared to a threshold value. If sales fall below this value the bit history will be defined as 1, otherwise it is 0. The threshold value is chosen such that 50% of all bits are 1 and 50% are 0. The actual history for the ketchup sales data found this way for the first year is:

$$History = 1101101100011111110110111000110001000010001000001110 \qquad 1$$

In general the history of the system will be denoted by H(t) ∈ {0, 1}. Once the system history is encoded in a binary string, we can use this to search for predictive strategies. A strategy is a complete collection of predictive rules. A rule is a prediction for a bit H(t) in the future, given that the previous M bits have particular values. M is the memory length on which the prediction is based.

For every point in time, the history has a particular value for the past M bits, the system state μ. These past bits can be interpreted as a binary number that is given by

$$\mu(t) = \sum_{k=0}^{M-1} 2^k H(t-k) \qquad 2$$

A strategy now is a binary map that assigns a bit to each of the $2^M$ possible values of μ:

$$a : (0,..., \mu,..., 2^M - 1) \to (0,1) \qquad 3$$

Hence for every given value of μ we look up the rule that applies according to the strategy. For example, for M = 3 a rule could be that after string 101 follows bit 1. Note that this rule always applies in Eq 1. Since there are $2^M$ possible values of μ, and each strategy can have two possible outcomes for each μ, the total number of strategies is $2^{2^M}$. For M = 3 all strategies can be checked systematically. However, the number of strategies quickly increases with memory length; for M = 4 there are 65536 possible strategies. Therefore the strategies are drawn randomly by assigning the $2^M$ rules (0 or 1) for each strategy by a random generator.

The success of a strategy is evaluated by evaluating its success rate of predicting the actual outcome. There are two possible ways to reward strategies, binary or continuous. In the binary reward scheme the strategy will collect a point when its prediction is correct, but otherwise it looses a point. The score of a strategy is the total of these points collected over time. However, to make a predictive tool we select a large pool of random strategies and evaluate their score to predict the actual history string $t_D$ time steps later. The delay time $t_D$ can be 1 time step (or week) as in the standard minority game [2], but here it has been varied up to 20 weeks. The score of the prediction at time t is then added to the score at the later time $t+t_D$. Let the score of strategy *a* be denoted by $P_a(t)$, then the points collected by the strategy are

$$P_a(t) = \sum_{t'=0}^{t-t_D} 2\delta_{a(t'),H(t'+t_D)} - 1 = P_a(t-1) + 2\delta_{a(t-t_D),H(t)} - 1 \qquad 4$$

This way each strategy scores (virtual) points against the actual history, but in a similar manner the whole collection of strategies can evaluated. After an initial learning period (typically 20 weeks) we start to use the current best strategy to predict the later history string. Hence the game continuously switches to the best strategy at every point in time. Thus, all retained strategies score virtual points that will be used later on to decide which strategy is the best at that point in time, and the game as a whole scores real points S, that will be used to assess its predictive power. At each point in time a prediction is made based *only* on past information. Therefore this is really a prediction, and not a fit to the data. This way of predicting sales history is actually a single agent game, where one agent holds a large pool of N strategies.

Alternatively, we can play a multi-agent game with N/2 agents, where each agent holds two strategies. Each strategy is now rewarded or punished proportional to a reward function R(t), which is derived from the market sales σ(t). The overall average sales under promotion <σ> is subtracted from the sales, so that the average reward is zero (it is a zero sum game), hence R(t) = σ(t) – <σ>. Each strategy *a* collects points $P_a$ according to

$$P_a(t) = P_a(t-1) + a(t-t_{pred})R(t) \qquad 5$$

where now $a(t) = \pm 1$ is the action of strategy *a* at time *t*. The prediction of the whole game is the sum of actions of all played strategies:

$$A(t+t_{pred}) = \sum_{i=1}^{N/2} a_i(t) \qquad 6$$

where $a_i(t)$ is the action of agent $i$ at time $t$. This sum is therefore a number between $-N/2$ and $+N/2$, which can be correlated to the actual reward function at the later time. This correlation is an objective measure of how well the game predicts the reward function, and is defined as

$$C = \frac{\sum_{t>t_{learn}} A(t)R(t)}{\sqrt{\sum_{t>t_{learn}} A(t)^2 \sum_{t>t_{learn}} R(t)^2}} \qquad 7$$

This second method allows us to predict sales peaks but the first method leads to less noisy predictions.

## 4. Results

We start with the single-agent game for ketchup data. In Figure 3 the average score over 100 games of memory M = 5 is shown, together with the standard deviation in the score, for a learning time of 20 weeks and a prediction period T = 100 weeks. In each game shown here the agent holds N = 5000 strategies. The predictive power of the game is quite insensitive to the number of strategies held. Indeed, if we do a random prediction over T time steps, we perform a random walk, picking up either –1 or +1 at each step. Thus the distribution of game scores S over all strategies is proportional to $\Psi(S) \sim \exp(-\frac{1}{2}S^2/T)$. Hence the number of strategies one needs to check to arrive at a particular maximum score $S_{max}$ is roughly $N \sim \exp(\frac{1}{2}S_{max}^2/T)$. Thus, the maximum score that can be expected from an ensemble of N strategies is given by

$$S_{max} = \sqrt{2T \ln N} \qquad 8$$

As a consistency check the time series has been split in five successive blocks of 20 weeks over which predictions were made. The variation over these five blocks is the same as the standard deviation that is found by averaging over many independent games, if we take into account that the relative spread in a 20 week period must be larger than for a 100 week period. This indicates two things. Firstly, all five blocks are statistically equal, and secondly averaging over time is the same as averaging over strategies.

For a one step prediction we find an average score of 55%, which is very much in line with the result that Johnson *et al* obtained for the Dollar-Yen exchange [13]. The unexpected result obtained here is that the best strategy at any given point in time shows an *anti-correlation* with the actual time series when the delay time is longer than 2 weeks. This points at an oscillation in the market. Though not fully understood, this result is just as useful as a predictability larger than 50%, as we just need to do the opposite as what the game would predict. The oscillation does not show up in a simple time auto-correlation function, see Figure 4, which is just statistical noise around zero.

To further investigate the market predictability, memory length has been varied. The average game score is presented in Figure 5 as function of the delay time. The score averaged over predictions from 4 up to 12 weeks, increases from about 43% at memory 3 to 48% at memory 7. To decide if the effect described is real or just a coincidental artefact of the data a number of tests were done. As a first test, the game theoretical analysis was applied to four series of random bit strings that were generated under the condition that 50% of the bits were 0, and 50% were 1, just as for the real data. For these random series the game prediction should be 50% accurate, since there is no correlation in the data [17]. Because a time series of 120 weeks is quite short, however, some noise in the prediction can be expected. Figure 6 shows these predictions for $M = 3$.

The results shown are consistent with a 50% accurate prediction, as should be expected. The results vary around the 50% line, and do not show a trend like the ketchup results do. Indeed, when long data series of 1200 points are generated, all predictions fall on the 50% line to within the expected 1% error. Taking a 100 game average of 100 independent random strings of 120 points, we found that the mean score per game is 49.1±0.1%, in line with the analytic result for these short strings [17]. Moreover, the prediction for random strings is independent of the delay time. For ketchup data, however, we find statistically significant deviations from this random string result. If we average the score from 3 to 12 weeks we find 43.1±2.4% score for $M = 3$ and 45.6±1.3% score for $M = 5$. The error bar is the spread in the score per delay. Thus, the deviation of the ketchup data is well outside the expected error bar for a range of delay times.

If this hidden correlation in the ketchup data is not based on coincidence, one might expect a similar behaviour for other products. Therefore the same analysis has been repeated for curry sauce, mayonnaise and barbecue sauce. The curry and mayonnaise data are available for exactly the same weeks as the ketchup sales data refers to; the barbecue sauce data comes from a later period. These results are statistically similar to the results for ketchup. The average over all markets is shown in Figure 7. This clearly indicates that the dip in the game theoretical prediction between 3 and 10 weeks is no coincidence, but is a genuine effect that may be used to improve promotion timings relative to competitors. The effect appears to be independent of the particular market.

For random strings the score as in Figure 6 is independent of the delay time. Hence, if the effect for the real markets is based on coincidence, all scores other than 50% (or 49.17% if we take into account that the strings are constrained to have half the bits at 1) should be independent of the forecast for other delay times. The ketchup, curry and mayonnaise histories in the same week are correlated because all sauces are often promoted in a single action. We correct for this correlation by taking out one data series. If we analyse the scores, we find that they are Gaussian distributed around a mean of 44.3%, with a spread of 2.9±0.3%. As we have 12 independent variables, the error in the mean is 0.9±0.1%. This means that the observed mean is 5 to 6 error bars away from the expected 49.17% for random data. The probability for this to occur by coincidence is less than $1:10^6$.

At this point the question arises, should we predict the full market or just the actions of the competitor? In the standard Minority Game the payoff is determined *including* the actions of the player himself. So what happens if we exclude one brand from the statistics, i.e. what happens if a brand manager looks at what all others do, disregarding the sales of his own brand? The underlying data for this analysis is the same as for Figure 7. However, for each market the game was applied to the market sales, excluding the sales of one brand. This excluded brand was then alternated over all brands. When the results for all markets are averaged we find the data shown in Figure 8. The puzzling result is obtained that all predictability disappears, apart from possibly the first week.

This result is puzzling, as this implies that the market behaves unpredictable if we leave out one player, whereas the market does have some predictability if we take all players into account. Without a careful analysis of the previous results, one might be tempted to conclude that if none of the parts of the market is predictable, the market as a whole must be unpredictable. The reason why the market is not predictable when one sales series is left out may be the following [18]. Cavangna [19] has demonstrated that it is not essential for the information vector μ to correspond to the recent history of the game. The main results do not change if the information vector is chosen randomly. The only thing that matters is that agents have a publicly available signal, which they could use to co-ordinate their actions. This signal could just as well be a measure of current sunspot activity and have nothing to do with the actual game. Hence, to reproduce the FMCG market correctly, it is essential that the actions are based on a signal that is common to all brand managers. If all brand managers base their actions on the sales of just the others, i.e. excluding their own sales, they all base their decisions on different information. In that case co-operativity between the actions could never arise, and therefore the market would be unpredictable. Predicting quasi-chaotic time series is not exclusive to the Minority Game. Neural networks are equally suitable for this purpose. With this technique it was found that there are time series that cannot be predicted [20], and moreover that learning and prediction are not necessarily related [21].

Although a decision to do or not to do a promotion is a yes or a no, the strongest impact on promotional sales comes from the highest sales peaks. Therefore it would be advantageous to be able to forecast the promotional sales itself, rather than the limited information above or below average. This is indeed possible if we play the Minority Game of a collection of agents against the sales of the market. To do this, rather than having one agent with 500 strategies, we use 250 (or in general N) agents, each holding 2 strategies. This game has been described in the previous section.

From the correlation as defined in Eq 7 the game score is found as S = C/2+50%. The mean score as function of the delay time, averaged over a 70-week period after a learning time of 50 weeks, is shown for four markets in Figure 9. The variation from one delay time to another and from one market to another seems to be larger than for single-agent games, particularly in the more interesting range from 6 to 10 weeks. Much of this noise can be attributed to the noisy reward function. Averaged over 20 weeks, however, the two methods give comparable spread between the markets. The trend observed for the single agent game is reproduced quantitatively by the multi-agent game. The best example is shown in Figure 10 for the barbecue sauce market, showing –A(t) and R(t). Note that the percentage of correct predictions of the sign is remarkably constant over a two-year period. This holds no guarantee for future, as the 7-week and 9-week forecasts are much less accurate. For other markets a larger scatter in predictability is obtained. The apparent predictability of the barbecue sauce market is possibly based on the selection of the right market and delay time, and is probably fortuitous.

## 5. Conclusions

The aim of this paper is to investigate if the Minority Game can be applied to improve the timing of promotional actions. The underlying assumptions for this approach are that promotions are more effective when fewer than average competitors do a promotion and that a promotion strategy to comply with this rule can be based on past sales data.

The first assumption was confirmed by analysing 1467 promotional actions for three products on the Dutch market (ketchup, mayonnaise and curry sauce) over the same 120-week period, both on an aggregated level and on retailer chain level. The second assumption has been tested by applying game theory to past sales data. This revealed that high or low competitor promotional pressure for actual ketchup, mayonnaise, curry sauce and barbecue sauce markets is to some extent predictable up to a forecast of some 10 weeks. Whereas a random guess would be right 50% of the time, a single-agent game theory can predict the market with a success rate of 56% for a 6 to 9 week forecast. This number is *the same* for all four mentioned fast moving consumer markets. For a multi-agent game a larger variability in the success rate is obtained in the relevant forecast period, but apart from this it agrees quantitatively with the single agent game.

Contrary to expectation, the actual market does the *opposite* of what game theory would predict. Even though this result is not yet understood, merely observing that this trend is present in the data will lead to exploitable trading benefits. The fact that the best strategy at each point in time is systematically below average in some 8 weeks time implies that there is a hidden correlation in the FMCG market. When any brand is left out of its market the sales history of that market becomes effectively random. This implies that each individual brand is unpredictable, and that the predictability of the market is a collective effect.

As a check on the results, randomly generated history strings were studied. This shows that the statistical variation of the game theory, applied to purely random time series of the same length, would lead to spurious forecast success rates between 45% and 55%. The effect observed on real data is therefore at the edge of the expected statistical variation. However, since four different markets lead to the same effect and indeed show a similar trend over several weeks, the odds are roughly 1:1,000,000 that this is based on coincidence.

## Figure Captions

**Figure 1** Total number of items of ketchup sold in the Netherlands over a 120 week period (top curve) and items sold under promotion (bottom curve). The dashed line gives the threshold amount above which promotional sale is deemed excessive.

**Figure 2** Ketchup sales under promotion for Calvé (red), Gouda's Glorie (blue), Heinz (green), and Remia (purple) anywhere in the Netherlands, as function of other ketchup promotions. The right hand figure gives the same plot, averaged for ketchup (red), mayonnaise (green) and curry sauce (blue) markets.

**Figure 3** Average score and spread in score as function of the delay time, obtained over 100 repeat games of memory $M = 5$. The score is presented as percentage of the maximum attainable score. The error bars give the spread over 100 games, the drawn line is a block average.

**Figure 4** Auto-correlation of the bit string history of ketchup data, showing no long time correlation. The scatter is distributed symmetrically around zero.

**Figure 5** Average game score as function of the delay time for games of memory $M = 3$ up to $M = 7$. The score of a random prediction is indicated by the dashed curve.

**Figure 6** Average game score for 4 random histories of the same size as available for the ketchup data, for $M = 3$. Error bars give the variation *between the series*.

**Figure 7** Average game score for ketchup data, curry sauce, mayonnaise, and barbecue sauce, for memory $M = 3$ based on a learning time of 25 weeks. Error bars give the estimated error in the mean based on the variation between the series. The line is a smooth fit to guide the eye.

**Figure 8** Game score for ketchup, curry sauce, and mayonnaise markets, excluding each brand in turn from the analysis. Bars give the estimated error in the mean based on variation between markets. Trends and variations are consistent with random series.

**Figure 9** Prediction score calculated from multi-agent game, averages for ketchup, curry, mayonnaise, and barbecue sauce markets over a 70-week average. The full curve is the trend observed for a single-agent game, see Figure 7.

**Figure 10** Prediction of 8-week forecast by taking the negative of the multi-agent game theory (grey) and actual barbecue sauce market data (black). The right-hand figure gives the percentage correct prediction over 20-week periods.

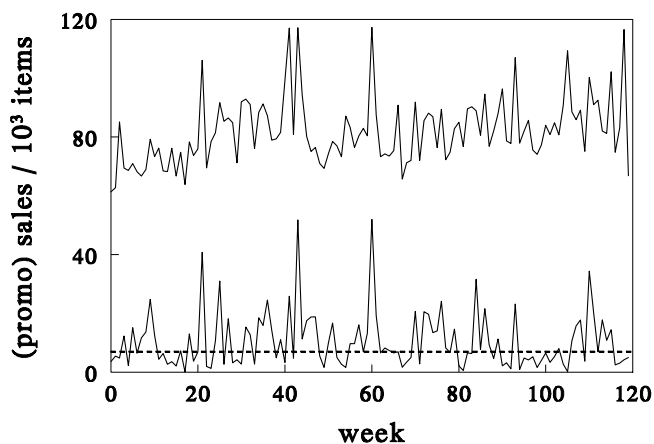

**Figure 1** Total number of items of ketchup sold in the Netherlands over a 120 week period (top curve) and items sold under promotion (bottom curve). The dashed line gives the threshold amount above which promotional sale is deemed excessive.

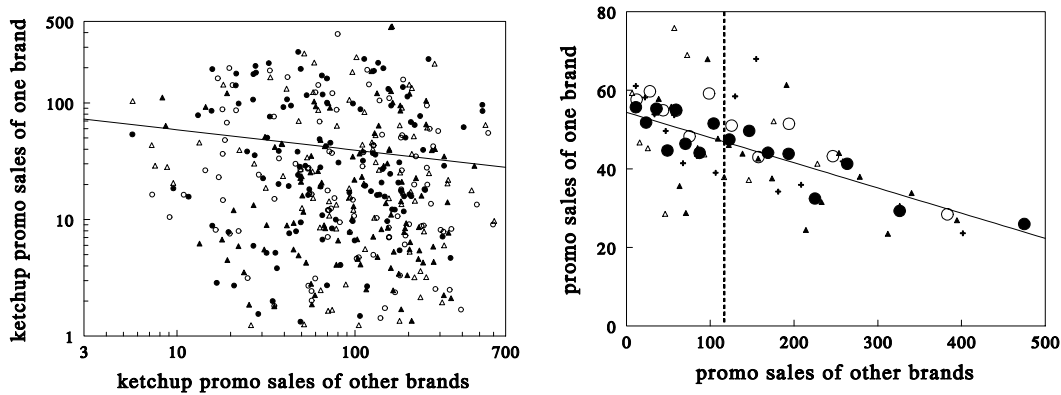

**Figure 2** Ketchup sales under promotion for Calvé (closed dots), Gouda's Glorie (open circles), Heinz (closed triangles), and Remia (open triangles) anywhere in the Netherlands, as function of other ketchup promotions. The right hand figure gives the same plot, averaged for ketchup (crosses), mayonnaise (open triangles) and curry sauce (closed triangles) markets. Large dots are average over all markets, large circles give the market average excluding Retailer Own Brand.

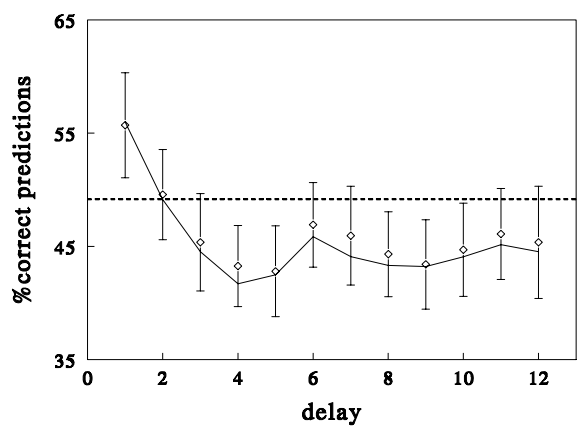

**Figure 3** Average score and spread in score as function of the delay time, obtained over 100 repeat games of memory M = 5. The score is presented as percentage of the maximum attainable score. The error bars give the spread over 100 games, the drawn line is a block average.

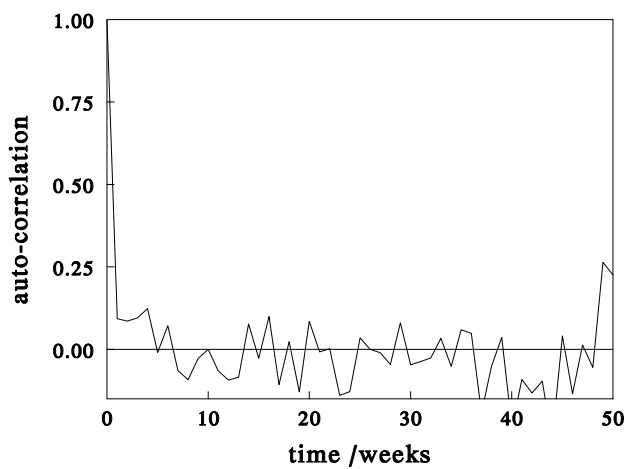

**Figure 4** Auto-correlation of the bit string history of ketchup data, showing no long time correlation. The scatter is distributed symmetrically around zero.

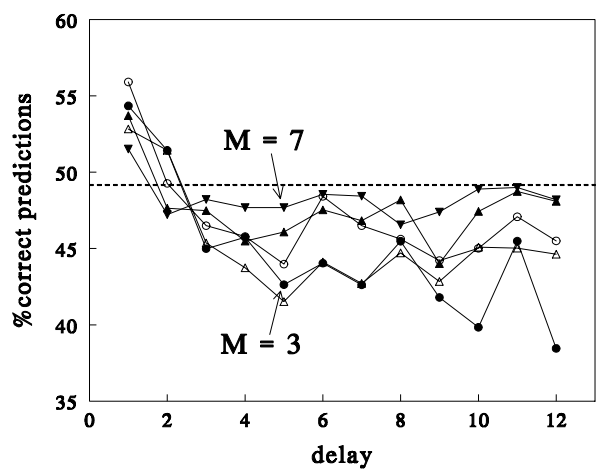

**Figure 5** Average game score as function of the delay time for games of memory M = 3 up to M = 7. The score of a random prediction is indicated by the dashed curve.

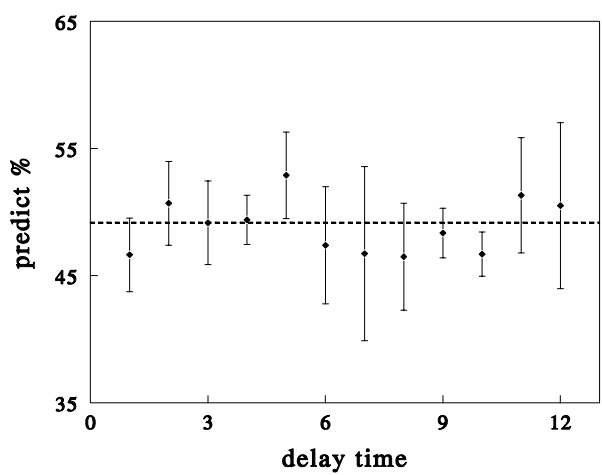

**Figure 6** Average game score for 4 random histories of the same size as available for the ketchup data, for M = 3. Error bars give the variation *between the series*.

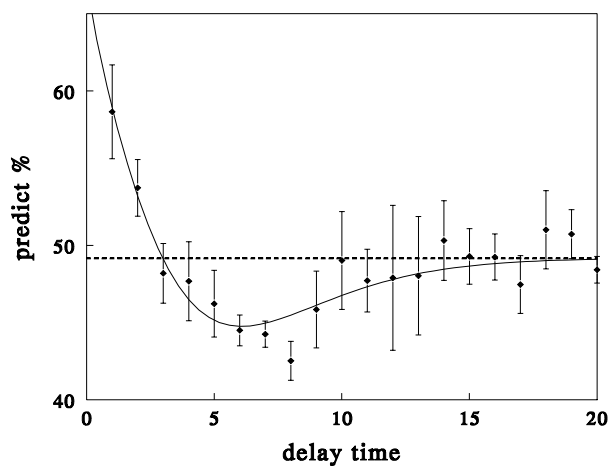

**Figure 7** Average game score for ketchup data, curry sauce, mayonnaise, and barbecue sauce, for memory M = 3 based on a learning time of 25 weeks. Error bars give the estimated error in the mean based on the variation between the series. The line is a smooth fit to guide the eye.

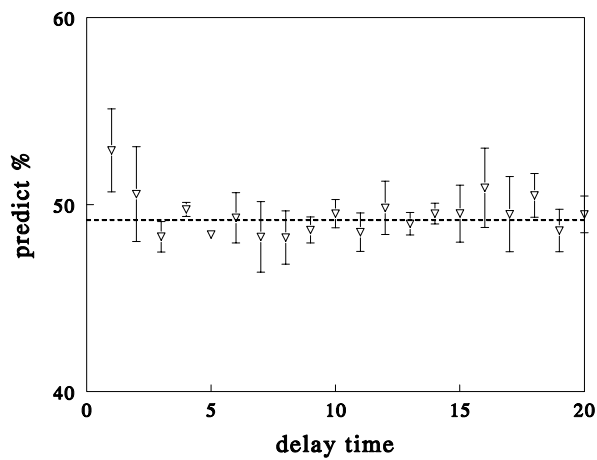

**Figure 8** Game score for ketchup, curry sauce, and mayonnaise markets, excluding each brand in turn from the analysis. Bars give the estimated error in the mean based on variation between markets. Trends and variations are consistent with random series.

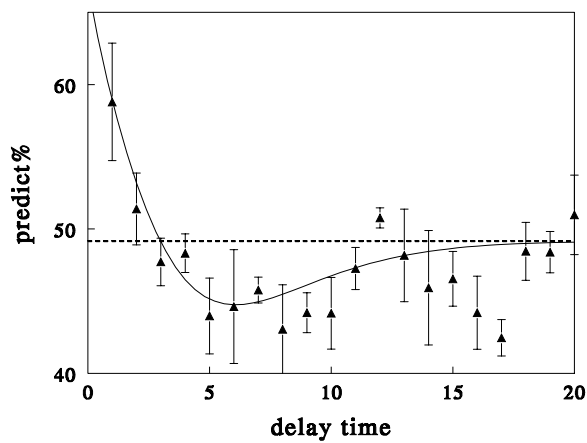

**Figure 9** Prediction score calculated from multi-agent game, averages for ketchup, curry, mayonnaise, and barbecue sauce markets over a 70-week average. The full curve is the trend observed for a single-agent game, see Figure 7.

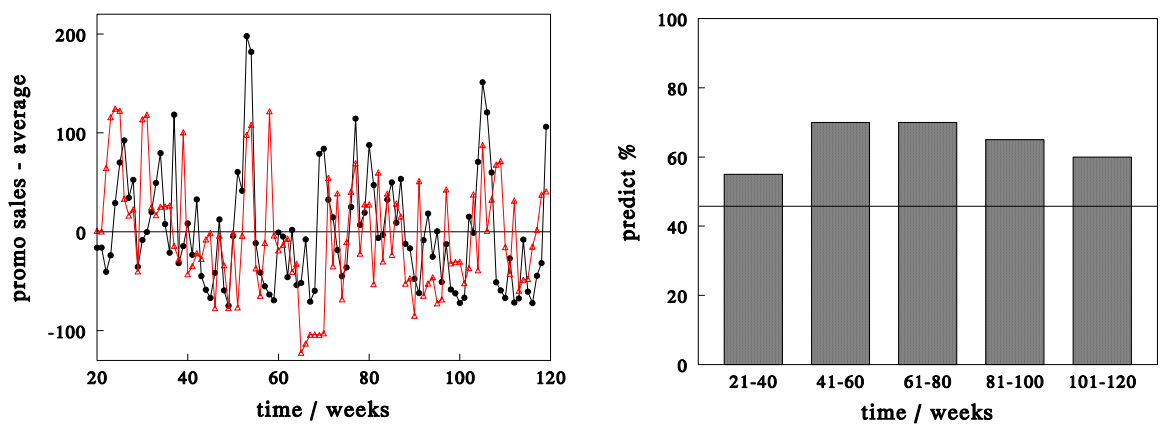

**Figure 10** Prediction of 8-week forecast by taking the negative of the multi-agent game theory (grey) and actual barbecue sauce market data (black). The right-hand figure gives the percentage correct prediction over 20-week periods.